\documentclass[aps,prl,twocolumn,groupedaddress,showpacs]{revtex4}
\usepackage{color}
\usepackage{graphicx}
\usepackage{ulem}
\usepackage{bm}

\begin{document}

% Use the \preprint command to place your local institutional report
% number in the upper righthand corner of the title page in preprint mode.
% Multiple \preprint commands are allowed.
% Use the 'preprintnumbers' class option to override journal defaults
% to display numbers if necessary
%\preprint{V 3.0}

%Title of paper

\title{A fluctuation relation for weakly ergodic aging systems}

% repeat the \author .. \affiliation  etc. as needed
% \email, \thanks, \homepage, \altaffiliation all apply to the current
% author. Explanatory text should go in the []'s, actual e-mail
% address or url should go in the {}'s for \email and \homepage.
% Please use the appropriate macro foreach each type of information

% \affiliation command applies to all authors since the last
% \affiliation command. The \affiliation command should follow the
% other information
% \affiliation can be followed by \email, \homepage, \thanks as well.
\author{A. Crisanti$^1$}%\email{andrea.crisanti@roma1.infn.it}
\author{M. Picco$^2$}%\email{picco@lpthe.jussieu.fr}
\author{F. Ritort$^3$}%\email{fritort@gmail.com}

%\thanks{}
%\altaffiliation{}
\affiliation{$^1$Dipartimento di Fisica, Universit\`a di Roma
             ``La Sapienza'', and \\ 
            Istituto dei Sistemi Complessi ISC-CNR, 
            P.le Aldo Moro 2, 00185 Roma, Italy}

\affiliation{$^2$CNRS, LPTHE, Universit\'e Pierre et Marie Curie,
                   UMR 7589, 4 place Jussieu, 75252 Paris cedex 05, France
	         }

%\email[To whom correspondence should be addressed: ]{fritort@gmail.com}
%\homepage[]{http://www.ffn.ub.es/ritort/}
\affiliation{$^3$Departament de F\'{\i}sica Fonamental, Facultat de F\'{\i}sica, Universitat de Barcelona, Diagonal 647, E-08028, Barcelona}
\affiliation{CIBER-BBN Center for Bioengineering, Biomaterials and Nanomedicine, Instituto de Salud Carlos III, Madrid}

\date{\today}

%%
%%  IMPORTANT
%%   Please update here with the version number to keep track of changes
%%     
\date{\$ V 13.3.0 2012/08/13 10:37 AC \$}

% History
% V 4.0.0  Version FR
% V 3.1.3   Version AC and MP to be sent to FR
 %                References must be fixed
 %                PACS must be fixed
 %                [ ]-sentences can be removed if space is needed
 % V 11.0.0  More detains in the argument
 % V12.1.0   Cleaned up last version Felix and few changes
 % V13.2.0   Last version ready for submission
 % V13.3.0   PACS

\begin{abstract}
\noindent
A fluctuation relation for aging systems is introduced, and verified by
extensive numerical simulations. It is based on the hypothesis of
  partial equilibration over phase space regions in a scenario of
  entropy-driven relaxation. The relation provides a
simple alternative method, amenable of experimental implementation, to
measure replica symmetry breaking parameters in aging systems. The
connection with the effective temperatures obtained from the
fluctuation-dissipation theorem is discussed.
\end{abstract}

% insert suggested PACS numbers in braces on next line
%05.40.-a Fluctuation phenomena, random processes, and Brownian motion
%05.70.Ln Nonequilibrium and irreversibile thermodynamics
%64.70.Pf Glass Transitions
%82.39.Pj Nucleic acids, DNA and RNA bases
% insert suggested keywords - APS authors don't need to do this
%\keywords{}

%\maketitle must follow title, authors, abstract, \pacs, and \keywords
\maketitle

%
%\section{Introduction.} 
Non-equilibrium systems are characterized by a net energy transfer (in
the form of work, heat, or mass) to the environment.  Aging systems
pertain to the category of weakly ergodic non-equilibrium systems
\cite{Bouchaud92,CugKur95} exhibiting slow relaxational dynamics and
strong history dependent effects where the fluctuation-dissipation
theorem (FDT) is violated \cite{CriRit03,MarPugRonVul08,HerOci02,Kometal11}. 
This has
led to the introduction of the concept of non-equilibrium or effective
temperature \cite{CugKurPel98}. Despite of the insight gained from
exactly solvable models and other conceptual attempts
(e.g. \cite{Palmer82,FraVir00}) we still lack a clear understanding of
the general picture describing aging systems. In contrast to stationary
systems, aging systems are described by two timescales: the waiting time
$t_w$ elapsed since the system was set in the non-equilibrium state and
the time $t > t_w$ at which measurements are taken. A characterization
of the full spectrum of fluctuations appears key for a
satisfactory understanding of the aging state.

Over the past years several results about energy fluctuations in
non-equilibrium states have been obtained under the heading of
fluctuation theorems (FTs) \cite{Jar08,Cro99}. FTs take slightly different mathematical
forms depending on the specific non-equilibrium context
\cite{Ritort08,Seifert12}. However all them share the same common
feature: they relate probabilities of absorbing and releasing a given
amount of energy under non-equilibrium conditions; they are useful in
small systems and short times where energy fluctuations can be directly
measured allowing us, for example, to extract free energies of kinetic molecular
  states from irreversible pulling experiments \cite{Ale12}.

In this work we present a theoretical derivation of a fluctuation
  relation in aging systems. The aging
fluctuation relation (AFR) is based on the hypothesis of partial
  equilibration in a scenario of entropy-driven slow relaxation. It can be written in terms of a phase-space contraction factor $x$
that bears resemblance to the order parameter $x(q)$ defined in the
context of  spin glasses \cite{SG}. The new relation is further supported by extensive 
numerical simulations. The existence of an AFR was
already suggested in Refs. [\onlinecite{CriRit04,Ritort04}] and
recently hinted at in a quench experiment of a gelatin droplet that
exhibits a sol-gel transition \cite{GomPetCil11}. In that reference heat distributions were measured and shown to satisfy a
fluctuation relation for a system in contact with two
baths at different temperatures. Yet it remains unclear whether the same
relation applies to other aging systems.   
In contrast, the new  relation we are proposing should be generally valid in aging
systems and is amenable to future
experimental verification in noise measurements of glass formers,
critical systems and small systems (e.g. single molecules).

%
%\section{The aging fluctuation theorem (AFT)} 
%label{sec:AFT}
{\sl The aging fluctuation relation (AFR).}  Consider a system
quenched at time $t=0$ from an high-temperature equilibrium state
down to temperature $T$ where it ages exhibiting
  slow relaxation and activated dynamics phenomena. Such aging state
is characterized by low entropy-production rate and 
loss of time-translational invariance \cite{CugDeaKur97}.
%, both facts leading to strong violations of the FDR \cite{CriRit03}.  
During the
aging process the system continuously exchanges energy with the
bath, but some relaxational events result in larger than typical
amounts of heat $Q$ released to the bath, leading to a net positive entropy production
$\langle\Delta S\rangle=\langle Q\rangle/T > 0$, where the brackets $\langle(\cdots)\rangle$ denote 
the average over dynamical histories.

To analyze the spectrum of heat or entropy production fluctuations in the aging state
we consider the Bochkov-Kuzovlev work fluctuation relation, 
 originally introduced as a generalization of the FDT \cite{BocKuz77}. 
After a time $t_w$, elapsed since the system was quenched, a constant external perturbation of 
strength $h$ coupled to an observable $A$ is applied to the system. This will
cause a change in  $A$ during its subsequent evolution.  
The entropy production during the time interval $[t_w, t]$ ($t\ge t_w$)
caused by the perturbation is equal to 
$\Delta  S_{t_w,t}=Q_{t_w,t}/T=h[A(t)-A(t_w)] / T=h\Delta A_{t_w,t}/T$. 
This  quantity has been termed exclusive work in \cite{CamTalHan11b} and
 satisfies a fluctuation relation if the system is equilibrated at
 $t_w$ (which is not the case here). $\Delta S_{t_w,t}$ is
 a fluctuating quantity changing upon repetition of the same
 experiment.
Moreover in the aging regime, where relaxation dynamics is 
ruled by the complex topological structure of the phase space, made of many almost degenerate metastable states,  $\Delta S_{t_w,t}$ displays a strong intermittent behavior.
This means that for fixed $t_w$ and $\Delta S$ the requirement $\Delta S_{t_w,t} = \Delta S$ defines
a very broad interval of times $t$ covering many, well separated, timescales.
%, consequence of the escape from metastable states.
%
In such a context a data analysis for fixed $t_w$ and $t$ may be
ambiguous since it may mix up processes with different timescales, as
noticed some years ago in the framework of turbulence
\cite{Aurelletal96, Aurelletal97}. To overcome this problem we define
$P_{t_w}(\Delta S)$ as the probability of observing the value $\Delta
S_{t_w,t}=\Delta S$ after  $t_w$. 
%This definition assumes
%that the maximum value of $t$ is large enough for collecting
%sufficient statistics for the relevant values of $\Delta S$.

In Ref. [\onlinecite{CriRit04}] a similar approach, based on the concept
of Inherent Structure, 
was used. The idea was to look for $\Delta S_{t_w,t}$ associated to the
first jump out of an Inherent Structure. While
this definition could in principle be used in a numerical simulation,  it is far less useful 
in a real experiment.  In contrast, the one proposed here is  well suited for real experiments.

Despite the fact that the perturbing field $h$ favors positive values of 
$\Delta S_{t_w,t}$, trajectories with negative values can be observed as
well.
We argue that, for long enough $t_w$, the probability of observing positive and negative 
values of $\Delta S$ satisfies the following fluctuation relation (AFR) 
\begin{equation}
\label{AFT}
  \log \left[ \frac{P_{t_w}(\Delta S)}{P_{t_w}(-\Delta S)} \right] = 
  x_{t_w}\frac{\Delta S}{k_{\rm B}},
\label{eqAFR}
\end{equation}
where $k_{\rm B}$ is the Boltzmann constant (equal to $1$ in the following).
If at $t_w$ the system is in equilibrium,  then $x_{t_w}=1$ and 
Eq. (\ref{eqAFR}) reduces to the Bochkov-Kuzovlev work fluctuation
relation \cite{BocKuz77,CamTalHan11b}. 

In aging systems, such as glasses,  the FDT is not violated for times $t-t_w \lesssim t_w$
after switching on the perturbation.  
This is because the degrees of freedom whose characteristic relaxation times are 
sufficiently smaller than $t_w$ have equilibrated. In the present context this
means that small enough energy transfers between the
system and the bath occur in equilibrium, and hence  $x_{t_w} \simeq 1$. 
This remains true as long as $|\Delta S|$ is smaller than
a cross-over value $ \Delta S^*$,
which sets the scale of the typical minimum energy transfer 
in processes that involve non-equilibrated degrees of freedom. 
For $|\Delta S| > \Delta S^*$   non-equilibrium energy exchange 
processes responsible of slow relaxation come into play, and  the equality 
$x_{t_w} = 1$ is violated. 
Not all degrees of freedom can contribute to the relaxation process, some of them being frozen
at $t_w$, and hence $x_{t_w} < 1$. 
As the system ages more degrees of freedom equilibrate at $t_w$, and hence 
$\Delta S^*$ increases with $t_w$.  In the limit $t_w\gg t_{\rm eq}$, where $t_{\rm eq}$ is the
 equilibration time,  all degrees of freedom 
have equilibrated and $x_{t_w}$ converges to 1 for all $\Delta S$.
We note that this may not be the case for mean-field models where metastable states have
infinite lifetime.

%========================

To justify Eq. (\ref{eqAFR}) we consider a system whose time evolution is ruled by the 
Langevin equation
\begin{equation}
\label{eq:lang}
 \frac{d \varphi}{d t} = -\frac{\delta}{\delta\varphi}  H(\varphi) + h +  \xi
\end{equation}
where $\varphi$ is an $N$-dimensional field, 
$\xi(t)$ a Gaussian white noise (thermal noise) of zero average  and correlation
$\langle \xi(t)\,\xi(t')\rangle = 2\,T\, \delta(t-t')$,  
$-\delta H(\varphi) /  \delta\varphi$
the force arising from the conservative energy $H(\varphi)$, and $h$ the  constant 
external field coupled to the macroscopic observable $\psi_t = \sum_i \varphi^i_t$, where $i$ 
denotes a site index.
Fluctuation relations derive from the behavior of the probability ${\cal P}$ of a trajectory
$\{\varphi_s\}_{s\in[t_w,t]}$ and its {\sl reverse} 
$\{\widetilde{\varphi}_s\}_{s\in[t_w,t]} \equiv \{\varphi_{t+t_w-s}\}_{s\in[t_w,t]}$.  
These can be easily computed using the path integral formalism, see e.g. \cite{Seifert05}:
\begin{equation}
e^{-\Delta S_{t_w,t} + \Delta S^{\rm eq}}\, {\cal P}[\{\varphi_s\}_{s\in[t_w,t]}]\, =
        {\cal P}[\{\widetilde{\varphi}_s\}_{s\in[t_w,t]}]
\end{equation}
where  $\Delta S_{t_w,t} = \beta h (\psi_{t} - \psi_{t_w} )$ and 
$\Delta S^{\rm eq} = \beta [ H(\varphi_t) - H(\varphi_{t_w}) ] = 
- \ln [P^{\rm eq}(\varphi_t) / P^{\rm eq}(\varphi_{t_w})]$.
The quantities in the exponent depend on the trajectory end-points only,  then summing over
all trajectories from $\varphi_{t_w}$ at $t_w$ to  $\varphi_t$ at time $t$,  and including 
normalized probability distributions $P_0(\varphi_{t_w})$ and 
$P_1(\widetilde{\varphi}_{t_w}) = P_1(\varphi_{t})$
for the initial and final states, we get 
\begin{equation}
e^{-\Delta S_{\rm tot} }\, P(\varphi_t,t | \varphi_{t_w}, t_w) P_0(\varphi_{t_w}) 
   = P(\varphi_{t_w},t | \varphi_{t}, t_w) P_1(\varphi_{t})
\end{equation}
where
$P(\varphi_t,t | \varphi_{t_w}, t_w) $ and $P(\varphi_{t_w},t | \varphi_{t}, t_w)$
are the conditional probabilities of the forward 
$\varphi_{t_w} \to \varphi_t$  and reverse  
$\varphi_{t} \to \varphi_{t_w}$ trajectories, respectively, and
$\Delta S_{\rm tot} = \Delta S_{t_w,t} - \Delta S^{\rm eq}  + \Delta S_{\rm b}$, with 
$\Delta S_{\rm b} =  - \ln [P_1(\varphi_t) / P_0(\varphi_{t_w})]$.
From this relation  the following identity follows
\begin{equation}
\label{eq:Pmds}
   P_{t_w,t}(-\Delta S) = 
     \left\langle 
         \delta\bigl( \Delta S_{t_w,t} - \Delta S \bigr)\,
         e^{-\Delta S_{\rm tot}} \right\rangle_{t_w,t}
\end{equation}
where the average is over the forward process $\varphi_{t_w} \to
\varphi_t$ with initial probability distribution
$P_0(\varphi_{t_w})$.

After the quench the system partially equilibrates inside independent
phase-space regions (that we will call {\sl cages}), from 
which it will escape only after a time $t-t_w \sim t_w$.
Therefore, when $t-t_w \ll t_w$ the system is in (partial) equilibrium with the thermal bath, 
$P_0(\varphi) = P_1(\varphi)\propto P^{\rm eq}(\varphi)$ so that
$\Delta S^{\rm eq} = \Delta S_b$, $\Delta S_{\rm tot} = \Delta S_{t_w,t}$, and from  (\ref{eq:Pmds}) one gets eq. (\ref{eqAFR}) with
$x_{t_w}=1$.

To study the opposite limit $t-t_w \gg t_w$, where the system can
access different cages, we observe that $\Delta S_{t_w,t}$ depends
only on the macroscopic variables $\psi_t$ and $\psi_{t_w}$, then the
average on the r.h.s of (\ref{eq:Pmds}) can be done by partial
classification, that is by  averaging first over all paths with given initial and final
states, and then over $\psi_t$ and $\psi_{t_w}$:
\begin{eqnarray}
\label{eq:Pmds1}
   P_{t_w,t}(-\Delta S) &=&  \int d\psi_{t_w} \int d\psi_{t}\,
         \delta\bigl( \Delta S_{t_w,t} - \Delta S \bigr) \nonumber\\
         &\phantom{=}& \hskip 10pt
      \times     \left\langle 
         e^{-\Delta S_{\rm tot}} \right\rangle_{\psi_{t_w}t_w; \psi_tt}.
\end{eqnarray}
where $\langle(\cdots)\rangle_{\psi_{t_w}t_w; \psi_tt}$ denote dynamical averages
restricted to those trajectories starting with $\psi_{t_w}$ at $t_w$
and ending with $\psi_{t}$ at $t$.
Assuming the system is partially equilibrated over cages, the probability $P(\varphi|\psi)$ of
a state $\varphi$ in a cage with fixed $\psi$ is proportional to
$P^{\rm eq}(\varphi)$ times the probability of having $\psi$ in a
cage.  Thus $ P(\varphi|\psi) \propto P^{\rm eq}(\varphi) \times
\Omega_{\rm cage}(\psi) / \Omega_{t_w}(\psi)$, where $\Omega_{\rm cage}(\psi)$ is the
number of states with $\psi$ inside the cage, divided by the total
number $\Omega_{t_w}(\psi)$ of accessible states, not necessarily in
the same cage, with $\psi$.  Under the hypothesis of partial
equilibrium $S_{\rm cage}(\psi) = \ln\Omega_{\rm cage}(\psi)$ is
the thermal equilibrium
entropy in the cage.
$\Omega_{t_w}(\psi)$ depends on system age since more we wait more
degrees of freedom relax.  $S_{t_w}(\psi) = \ln\Omega_{t_w}(\psi)$
is then smaller than the full thermodynamic entropy and converges to
it only for $t_w\gg t_{\rm eq}$.  Using this {\sl ansatz} for the PDF of
the initial and final states,  and
the relation $\partial S_{\rm cage}(\psi) / \partial \psi = \beta h$ 
together with the analogous 
$\partial S_{t_w}(\psi) / \partial \psi = x_{t_w} \beta h$, 
corrected through the coefficient 
$x_{t_w} < 1$ to account for the frozen degrees of
freedom, we have
$\Delta S^{\rm eq} - \Delta S_{\rm b} = 
           \beta h  (\psi_{t} - \psi_{t_w} ) - x_{t_w}\beta h  (\psi_{t} - \psi_{t_w} )
        =  (1-x_{t_w})  \Delta S_{t_w,t} $.
Inserting this form into (\ref{eq:Pmds1}) the AFR (\ref{eqAFR}) follows.
The coefficient $x_{t_w}$
measures the {\sl phase space contraction} due to frozen degrees of
freedom  at $t_w$.  Clearly $x_{t_w}\to 1$ as $t_w\gg t_{\rm eq}$.

%\section{Numerical tests of the AFR.} 
%\label{sec:num}
{\sl Numerical tests.}  We have tested the AFR (\ref{eqAFR}) through
Monte Carlo simulations of several model systems, but we report results
for only three of them.  The system, initially prepared in an high
temperature equilibrium state, is instantaneously quenched to a
temperature $T$ below the freezing transition temperature.  After $t_w$
a perturbation of small intensity, to ensure a good statistics for
trajectories with $\Delta S < 0$, is applied and the fluctuations
$\Delta A_{t_w,t}$ of the conjugated variable $A$ are recorded at fixed
time intervals $t-t_w$. The maximum recording time $t$ was taken much
larger than $t_w$ to ensure good statistics. The procedure was repeated
several times, $P_{t_w}(\Delta S)$ calculated from data binning and
Eq.(\ref{eqAFR}) tested to extract the value of $x_{t_w}$. To compare it
with the parameter $x = T / T_{\rm eff}(t_w)$ obtained from the FDT, the
fluctuation-dissipation (FD) plots in time-domain were also measured using the standard procedures \cite{CriRit03}. Both x parameters (the one derived from the AFR, Eq.(\ref{eqAFR}), and the one derived from FDT) asymptotically coincide under general assumptions, see \cite{Supp}.

%\subsection{The Models}
The first model  is the Random Orthogonal model (ROM)
defined by the Hamiltonian \cite{CriRit03}
\begin{equation}
\label{eq:Hsg}
 {\cal  H}=-\sum_{1\leq i<j\leq N}\,J_{ij}\sigma_i\sigma_j,
\end{equation}
where $\sigma_i=\pm 1$ are Ising spins and  $J_{ij}=J_{ji}$  quenched Gaussian variables 
of zero mean and variance $1/N$ satisfying
$\sum_k J_{ik} J_{kj} = 16\delta_{ij}$, with $J_{ii} = 0$.
The system is perturbed by a uniform 
magnetic field of strength $h$ conjugated to the total
magnetization $M(t) = \sum_i \sigma_i(t)$.

This model
describes structural glasses in the mode coupling theory (MCT) 
approximation and has a dynamical  MCT transition at $T_{\rm d} = 0.536$. 
Below $T_{\rm d}$ the system is dynamically confined into one of the many
(exponentially large in number) metastable states and cannot reach full 
equilibrium. The equilibrium transition occurs at the lower static (or Kauzmann) 
temperature $T_{\rm c} = 0.25$.
The low-temperature behavior of the model is described by a 
one-step replica symmetry (1RSB) breaking order parameter.
Typical signature of this is a two-slopes FD plot.

\begin{figure}[h] %  figure placement: here, top, bottom, or page
   \centering
   \includegraphics[width=3.3in]{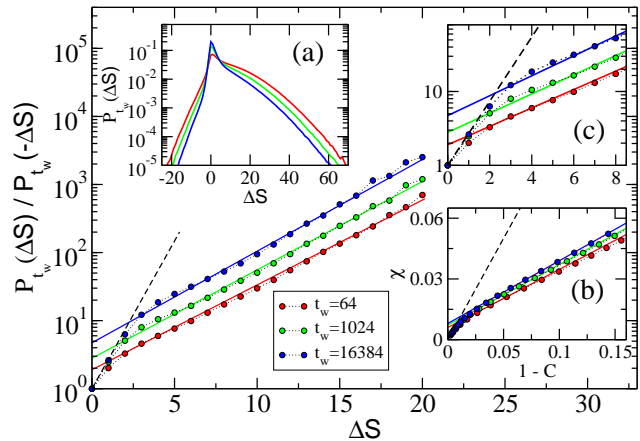} 
   \caption{{\bf Numerical test of the AFR in the ROM.} Main plot is a
     test of Eq.(\ref{eqAFR}) for the model with $N=1000$, $T=0.2$ and
     $h = 0.1$ at three values of $t_w$.  The dashed line corresponds
     to $x_{t_w}=1$ while full lines to $x_{t_w} = 0.271, 0.287,
     0.299$ for $t_w=64,1024,16384$ respectively. $(a)$
     $P_{t_w}(\Delta S)$ and $(b)$ FD plot.  The dashed and continuous
     lines shown in the FD plot have slopes equal to those shown in
     the main plot. $(c)$ Zoom of the region corresponding to
       $x_{t_w}=1$ (intra-cage relaxation).}
   \label{fig:PDSrr_ROM-N1000}
\end{figure}
Figure \ref{fig:PDSrr_ROM-N1000} shows results
for the ROM. Two regimes can be distinguished,
$x_{t_w} = 1$ for $|\Delta S| < \Delta S^*$ and $x_{t_w} < 1$ for
$|\Delta S| > \Delta S^*$ ($\Delta S^*\simeq 2$ for $t_w=1024$).  The values of $x_{t_w}$ agree quite
well with  $x = T / T_{\rm eff}$, extracted from
the FD plot [inset $(b)$].  

\begin{figure}[h] %  figure placement: here, top, bottom, or page
   \centering
   \includegraphics[width=3.3in]{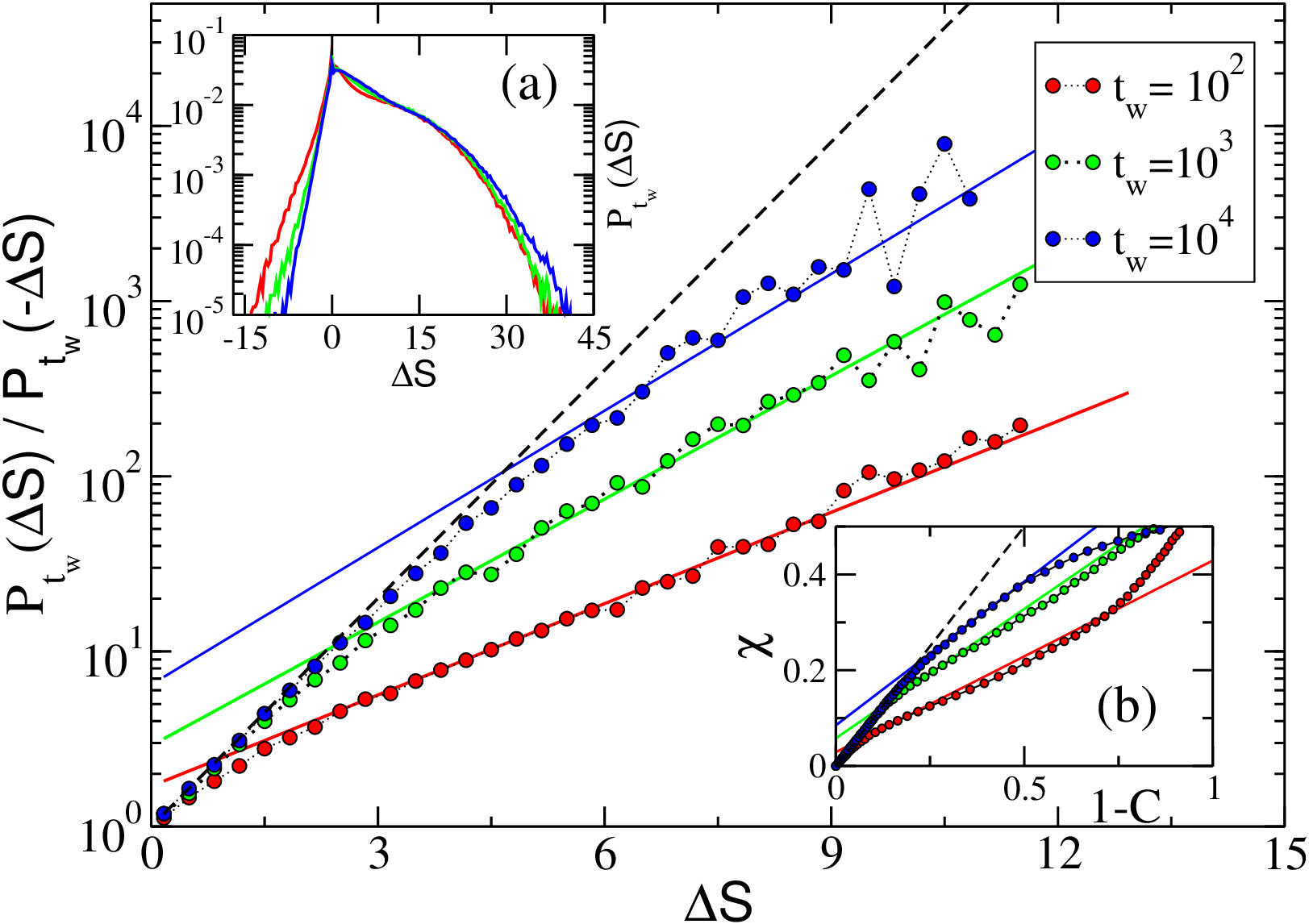} 
   \caption{{\bf Numerical test of the AFR in the BMLJ model.} Main
       plot is a test of Eq.(\ref{eqAFR}) for the model with parameters
       $N=500,V_0 = 0.1, |{\bm k}| = 7.25, T = 0.3$ at three values of $t_w$.  The dashed
       line corresponds to $x_{t_w}=1$  while full
       lines (red, green, blue) to $x_{t_w} = 0.40, 0.54, 0.60$
       for $t_w=10^2,10^3,10^4$ respectively. 
       $(a)$ $P_{t_w}(\Delta S)$ and $(b)$  FD plot.
       The dashed and  continuous lines (red, green, blue)  in the FD plot have slopes 
         equal to those shown in the main plot.
         }
   \label{fig:PDSr_BMLJ-N500}
\end{figure}

% BMLJ
As a more realistic system we have studied a $80:20$ binary mixture of type $A$ and $B$ particles 
interacting via a Lennard-Jones pair potential (BMLJ):
\begin{equation}
 V_{\alpha\beta}(r) = 4\,\epsilon_{\alpha\beta} \left[
       \left(\frac{\sigma_{\alpha\beta}}{r}\right)^{12}
              -
       \left(\frac{\sigma_{\alpha\beta}}{r}\right)^{6}
                                             \right] 
\end{equation}
where $\alpha, \beta = A, B$,
$r$ is the distance between the two particles and the parameters
  $\sigma_{\alpha\beta},\epsilon_{\alpha\beta}$ stand for the effective
  diameters and well depths between species $\alpha,\beta$. The parameters
for length and energy measured in units of $\sigma_{AA}$ 
and $\epsilon_{AA}$  are
$\epsilon_{BB} = 0.5$, $\epsilon_{AB} = 1.5$, 
$\sigma_{BB}=0.88$ and $\sigma_{AB} = 0.80$, 
and are taken to prevent crystallization \cite{KobAnd94}.
With this choice a system of reduced density $\rho = 1.2$
exhibits a glass transition
well described by the MCT at the critical temperature $T_{\rm MCT}  \simeq 0.435$.
The study of the AFR was done by adding at time $t_w$ an external
potential of the form $ V_0 \sum_j \epsilon_j \cos(\bm{k \cdot r})$,
where $V_0 < T$ and $\epsilon_j$ are i.i.d. (quenched) random
variables equal to $\pm 1$ with equal probability, and recording the
conjugated observable $A_k(t) = \sum_j \epsilon_j \exp[i\bm{k} \cdot
  \bm{r}_j(t)]$ \cite{BarKob99} . Results for the AFR are shown,
  and compared with standard FD plots, in
  Fig~\ref{fig:PDSr_BMLJ-N500}. Also in this case the agreement
  between the x extracted from AFR and that from FD plot is rather
  good.  Interestingly curves in the AFR plot do not exhibit the
  bending that can be seen in FD plots. Such bending is due to
  finite-size effects that for short $t_w$ produce an equilibration of
  the system, and a depletion of the statistics of rare events for
  longer $t_w$. The plot then bends upward for short $t_w$ and downward for
  longer $t_w$.
% end BMLJ

\begin{figure}[ht] %  figure placement: here, top, bottom, or page
   \centering
   \includegraphics[width=3.3in]{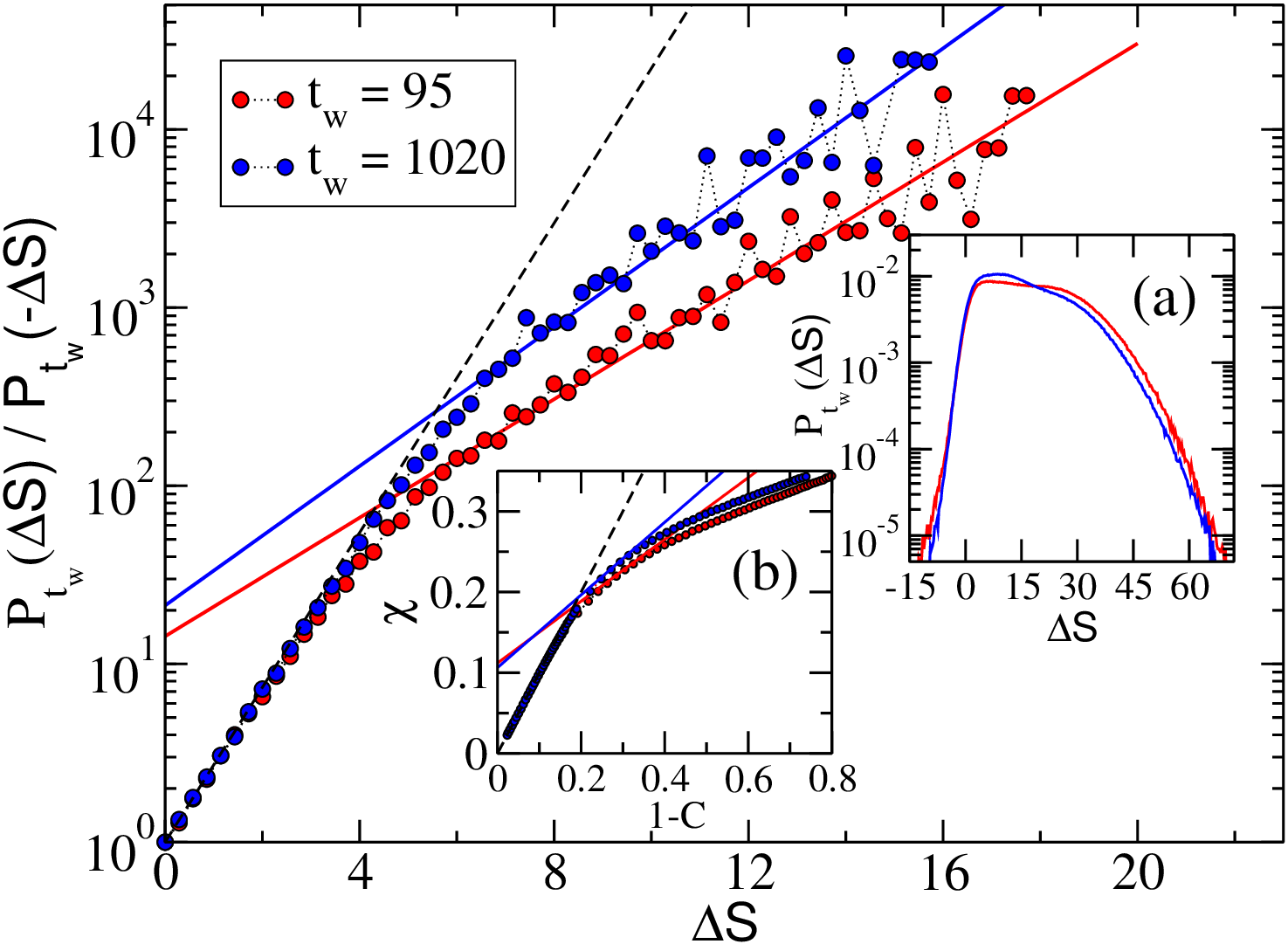} 
   \caption{{\bf Numerical test of the AFR in the $3D$ $\pm J$ EA model.} Main
       plot is a test of Eq.(\ref{eqAFR}) for the model with parameters $L=16$, $T=0.7$ and $h = 0.1$
       at two values of $t_w$.  The dashed
       line corresponds to $x_{t_w}=1$  while full
       lines (red,blue) to $x_{t_w} =  0.383, 0.450$ 
       for $t_w=95,1020$ respectively. $(a)$  $P_{t_w}(\Delta S)$ and $(b)$ FD plot.
       The dashed and continuous lines (red,blue)  in the FD plot 
        have slopes  equal to those shown in the main plot.
        }
   \label{fig:PDSrr_EA-L16}
\end{figure}

% EA
Both above systems are described by a two-steps relaxation, or 1RSB, scenario. 
As last example we have considered the $3$-dimensional  $\pm J$
Edwards-Anderson model
($\pm J$-EA) defined by the Hamiltonian (\ref{eq:Hsg}), but  with
$J_{ij}$ randomly chosen equal to $\pm 1$ if the sites $j$ and $i$ are 
nearest-neighbors on a cubic $3$-dimensional lattice, and zero otherwise.
Numerical investigation indicates that  below $T \simeq 1.14$ there is a spin-glass
phase described by a continuous-step relaxation, or Full-RSB, scenario. 
Typical signature of this is a continuous-slope FD plot. 
Numerical tests of Eq.(\ref{eqAFR}) for the $\pm J$-EA at two different $t_w$ are shown in
Fig.\ref{fig:PDSrr_EA-L16}, together with the FD plot.
The AFR is well verified also in this model. However, as shown in
  inset (b) and in contrast to the previous 1RSB models, 
the phase contraction coefficient $x_{t_w}$ agrees with $T/T_{\rm eff}$ only at the early stage,
where the FD plot depart from equilibrium. 
This is in agreement with the aforementioned argument, according 
to which $x_{t_w}$ gives information on the phase-space partition
at $t_w$.
% end EA

{\sl Discussion.}  Summarizing, the AFR (\ref{eqAFR}) shows a promising
route to experimentally test the {\it partial equilibrium-entropy
  driven} scenario  in slowly relaxing
systems from noise measurements. A theoretical derivation of the AFR was given and its validity
verified by extended numerical experiments. The connection between the
values of $x_{t_w}$ extracted from the AFR and FD plots was
shown. Remarkably enough, and in contrast to FD plots, extracting the
value of $x_{t_w}$ does not require measuring aging correlation
functions. By only measuring the statistics of $\Delta S$ to an externally applied perturbation, $P_{t_w}(\Delta S)$, we can
test the validity of Eq.(\ref{eqAFR}) to extract the value of $x_{t_w}$.
We emphasize that in order to test Eq.(\ref{eqAFR}) statistical events with $\Delta S<0$ must be observed. Since the average
value of $\Delta S$ continuously increases with $t$, only rare events
with $\Delta S<0$ give full meaning to the AFR. A similar situation is
encountered in the Gallavotti-Cohen theorem for steady state systems
\cite{GalCoh95}. Eq.(\ref{eqAFR}) is ready to be employed in
mesoscopic systems (e.g. magneto-conductance fluctuations in spin
  glasses and electron glasses \cite{Jar98,Car08}) and single molecule
  experiments. The latter include molecular systems exhibiting slow
  folding due to disorder and frustration (e.g. RNA) or slow binding
  kinetics (e.g. peptides or proteins binding DNA). Ultimately, small systems may provide a direct access to
experimentally measure the always elusive spin-glass order parameter.

\begin{acknowledgments}
AC thanks the LPTHE where part of this work was done. FR acknowledges
support from HFSP Grant No. RGP55-2008, ICREA Academia 2008 and Spanish Research Council Grant. No. FIS2010-19342.
\end{acknowledgments}

\bibliographystyle{unsrt}

\end{document}